\documentclass[conference]{IEEEtran}
\usepackage[utf8]{inputenc}
\usepackage[T1]{fontenc}

\usepackage{listings}
\usepackage{xcolor}
\usepackage{float}
\usepackage{wrapfig}
\usepackage[colorinlistoftodos]{todonotes}
\usepackage{tabularx}
\usepackage{ragged2e}
\usepackage{makecell}
\usepackage{rotating}

\usepackage{algorithm}
\usepackage{algpseudocode}
\usepackage{booktabs}   
\usepackage{pifont}    
\usepackage{fontawesome5}
\usepackage{multirow}

\definecolor{darkred}{RGB}{139, 0, 0} 

\usepackage{tikz}
\usetikzlibrary{shapes.geometric, arrows, positioning, shadows, arrows.meta}

\definecolor{challengecolor}{rgb}{0.8, 0.2, 0.2}
\newcommand{\challengecircled}[1]{\tikz[baseline=(myanchor.base)] 
\node[circle,fill=challengecolor,draw=challengecolor,text=white,inner sep=1pt] 
(myanchor) {\footnotesize #1};}

\usepackage{siunitx}
\usepackage{subcaption}
\usepackage{wrapfig}

\usepackage{pgfplots}
\pgfplotsset{compat=1.17}

\definecolor{cmarkgreen}{RGB}{0, 153, 0}  
\definecolor{xmarkred}{RGB}{204, 0, 0}  

\usepackage[table]{xcolor}
\usepackage{pifont}

\definecolor{tabgreen}{HTML}{E8F5E9}
\definecolor{tabred}{HTML}{FDECEA}

\definecolor{cmarkgreen}{RGB}{0,120,0}
\definecolor{xmarkred}{RGB}{170,0,0}

\newcommand{\cmark}{\textcolor{cmarkgreen}{\ding{51}}}
\newcommand{\xmark}{\textcolor{xmarkred}{\ding{55}}}

\newcommand{\yes}{\cellcolor{tabgreen}\cmark}
\newcommand{\no}{\cellcolor{tabred}\xmark}

\usepackage[hidelinks]{hyperref}
\hypersetup{
    colorlinks=true,
    linkcolor=blue,
    citecolor=blue,
    urlcolor=blue
}

\usepackage[shortlabels, inline]{enumitem}
\usepackage{xspace}
\usepackage{csquotes}

\usepackage{color, xcolor}

\definecolor{darkpurple}{RGB}{75,0,130}

\newcommand{\sepehr}[1]{{\color{black}#1}}

\newcommand{\subhead}[1]{\par\noindent\textbf{#1: }\ignorespaces}
\newcommand{\eg}{\textit{e.g.,}\xspace}
\newcommand{\ie}{\textit{i.e.}\xspace}

\newcommand{\mitre}{MITRE ATT\&CK\xspace}

\newcommand{\sol}{\textsc{Arms}\xspace}

\usepackage[most]{tcolorbox}
\usepackage{cite}
\usepackage{xurl}

\newcolumntype{Y}{>{\centering\arraybackslash}X}
\newcolumntype{C}[1]{>{\centering\arraybackslash}p{#1}}

\definecolor{autosigmaYellow}{HTML}{FFA33C}

\definecolor{rqtagcolor}{RGB}{0.1, 0.6, 0.4} 

\usepackage[most]{tcolorbox}
\newlength{\rqtagwidth}
\newlength{\rqtaghalf}
\newcommand{\rqline}[2]{%
  \settowidth{\rqtagwidth}{\bfseries\footnotesize #1}%
  \addtolength{\rqtagwidth}{8pt}
  \setlength{\rqtaghalf}{.5\rqtagwidth}%

  \begin{tcolorbox}[
    enhanced, sharp corners,
    colback=white, colframe=black, boxrule=0.5pt,
    left=\rqtagwidth, right=2mm, top=1mm, bottom=1mm,
    leftrule=0pt,
    overlay={
      \fill[rqtagcolor] (frame.north west) rectangle ([xshift=\rqtagwidth]frame.south west);
      \node[anchor=center, text=white, font=\bfseries\footnotesize] at ([xshift=\rqtaghalf]frame.west) {#1};
    },
  ]
  \textbf{\small #2}
  \end{tcolorbox}%
}

\begin{document}

\title{Automated Rule Mutation for Emerging Attack Variants using Infrastructure-Aware Technique Proximity}

\author{ 
    \IEEEauthorblockN{
    Sepehr Ghaffarzadegan\IEEEauthorrefmark{5},
    Boubakr Nour\IEEEauthorrefmark{1},
    Makan Pourzandi\IEEEauthorrefmark{1},
    Mourad Debbabi\IEEEauthorrefmark{5}, and
    Chadi Assi\IEEEauthorrefmark{5}
    }
    
    \IEEEauthorblockA{\IEEEauthorrefmark{5}Concordia University, Canada \qquad 
    \IEEEauthorrefmark{1}Ericsson Security Research, Canada} }

\maketitle

\begin{abstract}
    Detection rules are typically crafted based on specific observations of an attack. While these rules are effective for particular attack instances, their coverage diminishes when adversaries modify their techniques, such as by employing alternative execution paths or adapting to different telemetry environments while achieving the same objective. 
    To address this challenge, we introduce \sol, an \underline{a}utomated \underline{r}ule \underline{m}utation \underline{s}olution that 
    generates variant detection rules. 
    \sepehr{\sol defines a novel three-dimensional proximity space to identify techniques related to the original attack that capture structural relations, functional similarity, and environmental constraints.} 
    \sepehr{Rather than relying on ungrounded generation, \sol leverages verifiable attack execution commands, which are extracted via semantic alignment with the Atomic Red Team repository, to construct a context-aware Retrieval-Augmented Generation (RAG) pipeline.}
    %
    We evaluated \sol using verified, public SigmaHQ rules and successfully generated 
    variants, demonstrating \sol's capability to expand detection coverage.
    We further evaluated \sol across multiple research questions, showing the effectiveness of each proximity dimension and infrastructure-aware rule customization. 
    \sepehr{Nearly all generated variant rules successfully compiled into SIEM queries, and $\approx92\%$ of the newly generated rules detect the original behavior despite attackers' technique variations.}
    
    \faYoutube~ PoC demo: \url{https://youtu.be/QbGtuGGEnmQ}
\end{abstract}

\begin{IEEEkeywords}
    Cyber Threat Intelligence, Security Automation, Detection Rule Generation, Detection Rule Mutation
\end{IEEEkeywords}

\IEEEpeerreviewmaketitle

\section{Introduction}
\label{sec:intro}
Detection engineering has become an enduring and escalating challenge for modern organizational security~\cite{alaeifar2024current}. As organizations try to strengthen their defense by methods such as monitoring and rule-based detection, threat actors increasingly respond by changing how they can achieve the same goal through tool substitution, technique switching, and cross-platform execution changes~\cite{nour2025threatify}. Recent advancements in automation and artificial intelligence, including the widespread availability of Generative AI-based tools \cite{openai2024gpt4technicalreport}, have further diminished the cost of rapidly generating new scripts and commands, thereby accelerating the rate at which attackers can iterate through behavioral variants.

In practice, detection rules are implemented using diverse technologies. Security teams develop detection rules as SIEM queries (\eg Splunk SPL, Elastic/ES|QL, Microsoft Sentinel KQL), endpoint detections, or signature-like rules associated with specific telemetry sources \cite{shen2024decodingmitreengenuityattck}. This diversity and variety create an operational challenge. 
\sepehr{Even though the detection logic remains the same, the cost of translating rules from one environment to another is often very high,}
and even small environmental differences can make a correct rule unusable. The Sigma rule format \cite{roth2017sigma} addresses part of this challenge by offering a vendor-agnostic representation for detection logic, facilitating the sharing and translation of rules across different platforms and their mapping to the \mitre framework \cite{strom2018mitre}.

\smallskip
\subhead{Challenges}
Gathering Sigma rules is not tied to a single method or a single source of evidence. In real deployments, a Sigma rule may be 
\begin{enumerate*}[(i)]
    \item handcrafted by security practitioners, 
    \item adopted from shared repositories, 
    \item exchanged across organizations, or
    \item produced by automated \sepehr{rule generation} tools.
\end{enumerate*}
Regardless of how the rule is obtained, it is typically static~\cite{canotescape2023}. It codifies assumptions about a particular attack instance and about the telemetry for that particular attack. This brittleness becomes visible when an attacker changes techniques while preserving the intent. For example, a rule written for a PowerShell-based procedure will fail once the same goal is achieved via an alternate interpreter, a different execution primitive, or a platform shift. At the same time, directly deploying external or generic rules often triggers operational friction because enterprises differ in their OS mix, logging coverage, schemas, and legitimate administrative activity~\cite{maxam2024interview}. Without an awareness of the infrastructure, rules may become redundant, inapplicable, and generate excessive noise.

\smallskip
\subhead{Limitation of existing solutions}
Previous research has investigated methods to reduce the cost of generating initial detections, including approaches that apply Natural Language Processing (NLP) and Large Language Models (LLMs) to generate detection rule candidates from unstructured sources \cite{schwartz2025llmcloudhunter,xu2024intelex}. 
Other attempts tend to address the operational cost of large rule sets \cite{rulegenie2025} by optimizing and consolidating rule sets to reduce redundancy and improve efficiency.
As shown in Table~\ref{tab:related_work}, these approaches do not address fundamental challenges frequently encountered in detection operations: \challengecircled{1} maintaining coverage when attackers alter their techniques, and \challengecircled{2} ensuring that the rules remain valid and compatible with the target environment. 
Although evasion-aware rule hardening \cite{canotescape2023} can enhance robustness against bypass patterns, \challengecircled{3} it does not address technique substitution and behavioral drift outside the original rule's scope.

\begin{table}[!t]
    \centering
    \caption{Comparison with representative state-of-the-art solutions.}
    \label{tab:related_work}
    \renewcommand{\arraystretch}{1.2}
    \setlength{\tabcolsep}{5pt}
    \arrayrulecolor{gray!90}
    \resizebox{\linewidth}{!}{
    \begin{tabular}{p{2cm}lccc}
    \hline
    \rowcolor{gray!15}
    \centering\textbf{Category} & \textbf{Ref.}
    & \makecell[c]{\textbf{Variant}\\\textbf{Discovery}}
    & \makecell[c]{\textbf{Actionable}\\\textbf{Logic}}
    & \makecell[c]{\textbf{Infrastructure}\\\textbf{Awareness}} \\
    \hline

    \multirow{5}{2.1cm}{\centering Static Rule\\Generation}
    & \textsc{Falcon}~\cite{falcon2025} & \no & \yes & \no \\
    & \textsc{SigmaGen}~\cite{nightwolf2025} & \no & \yes & \no \\
    & \textsc{LlmCloudHunter}~\cite{schwartz2025llmcloudhunter} & \no & \yes & \no \\
    & \textsc{RulePilot}~\cite{wang2025rulepilotllmpoweredagentsecurity} & \no & \yes & \no \\
    & \textsc{ThreatPilot}~\cite{xu2024intelex} & \no & \yes & \no \\  
    \hline

    \makecell[c]{Rule\\Optimization}
     & \textsc{RuleGenie}~\cite{rulegenie2025} & \no & \yes & \yes \\
    \hline

    \makecell[c]{Rule\\Mutation}
     & \sol & \yes & \yes & \yes \\
    \hline
    \end{tabular}}
\end{table}

\smallskip
\subhead{Contributions}
To address these challenges, we propose \sol, an \underline{a}utomated \underline{r}ule \underline{m}utation \underline{s}olution designed to generate resilient, infrastructure-aware detection rule variants. Unlike previous work~\cite{schwartz2025llmcloudhunter,xu2024intelex} that focuses on static rule generation, \sol emphasizes the proactive mutation of detection rules to anticipate emerging threat variants while ensuring validity and compatibility with the target environment.
In this paper, we make the following contributions:
\begin{itemize}
    \item 
    To address challenge \challengecircled{1}, \sol defines a novel three-dimensional space to identify attack variants based on the logic of adversarial equivalence. \sol identifies technique proximity based on structure (\eg graph-based relationships), function (\eg semantic similarity), and environment (\eg shared prerequisites and exploitable vulnerabilities via CAPEC~\cite{mitre2024capec} mappings).
    
    \item 
    To address challenge \challengecircled{2}, \sol implements a mechanism to filter and customize detection rules according to an enterprise's specific infrastructure. By utilizing a structured infrastructure profile of organizational assets, \sol ensures that generated rules are optimized and relevant to the environment in which they are deployed.
    
    \item 
    To address challenge \challengecircled{3}, \sol utilizes Retrieval-Augmented Generation (RAG)~\cite{lewis2020retrieval} and Chain-of-Thought (CoT) reasoning~\cite{wei2022chain} to convert detection logic across different technologies, such as converting Windows-centric Sysmon logic into equivalent Linux auditd or cloud-native signatures.
    
    \item 
    We established a robust methodology to assess generated variant rules across three critical research questions: variant fidelity, rule correctness (including syntactic and semantic validity), and infrastructure relevance. 
\end{itemize}


\section{Motivation and Problem Statement}
\label{sec:motivation}

\subhead{Motivation} 
Security teams rely on Sigma rules to support detection and to address diverse operational contexts. However, the manual creation of Sigma rules is a significant bottleneck. Traditional manual workflows are labor-intensive, error-prone, and require deep domain expertise, making them difficult to scale in the face of a rapidly expanding threat landscape.
Despite the growth of open-source repositories like SigmaHQ~\cite{roth2017sigma}, these collections are inherently reactive. 
While attackers are dynamic and evolutionary, rules remain static. A rule designed to catch a specific PowerShell-based credential dumping technique becomes obsolete the moment an adversary switches to a Mac OS host or a direct API implementation.
Furthermore, copying and pasting rules generated by automated solutions introduces significant operational friction. Most automated solutions generate rules based on the specific telemetry of the organization that was originally attacked. However, every enterprise possesses a unique digital fingerprint, including specific log schemas, network topologies, and legitimate administrative behaviors.
Importing generic or external rules without environmental context leads to the deployment of redundant rules that target non-existent assets and the generation of excessive false-positive alerts.

\begin{figure}[!t]
    \centering
    \includegraphics[width=\linewidth]{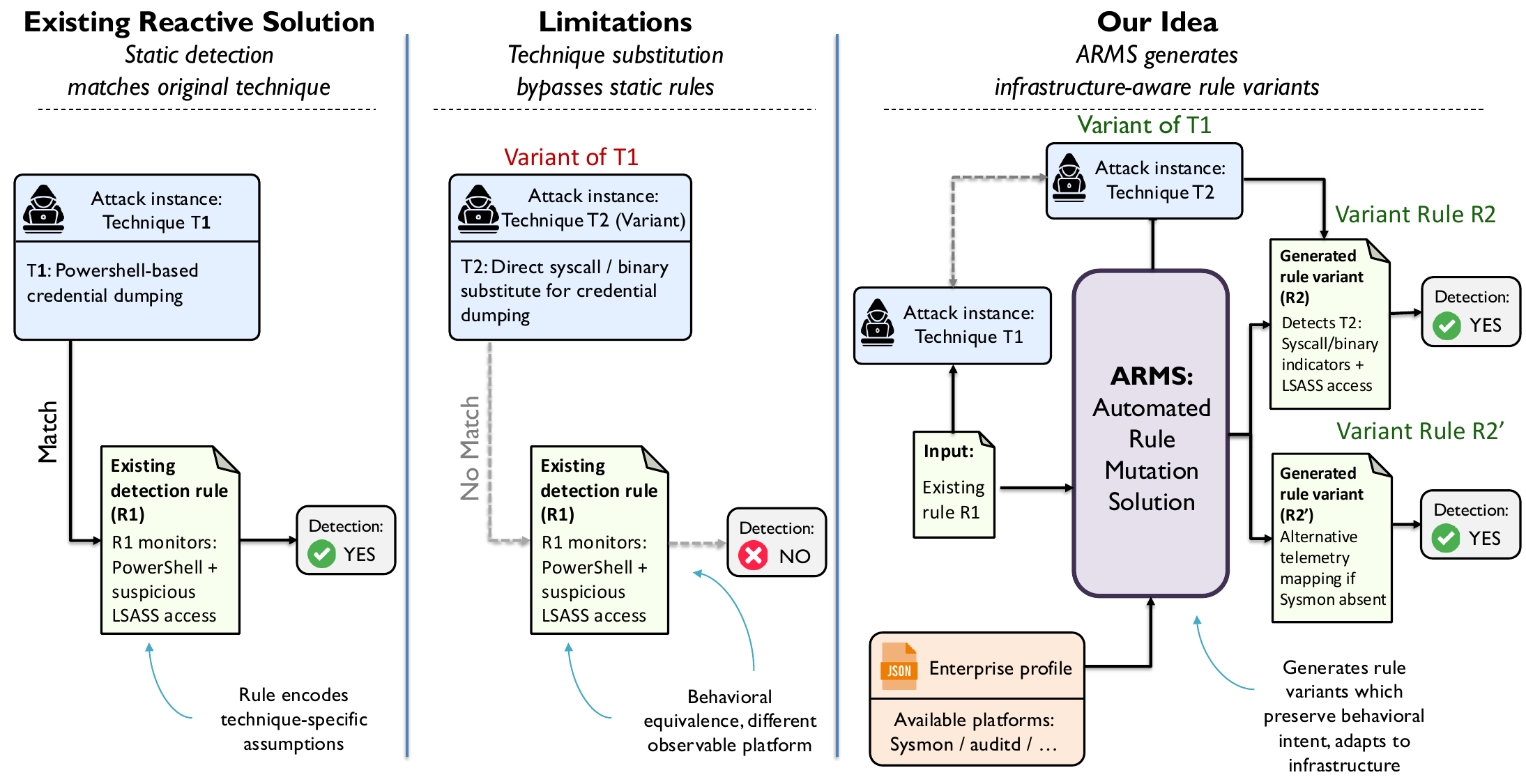}
    \caption{Illustration of the problem statement and motivation behind \sol.}
    \label{fig:motivating_example}
\end{figure}

\smallskip
\subhead{Problem Statement} 
Current research in automated detection engineering primarily focuses on the direct generation of structured and static rules \cite{xu2024intelex,schwartz2025llmcloudhunter,sigmagen2025,rulegenie2025,falcon2025}. As shown in Figure~\ref{fig:motivating_example} (left and middle sections), this approach is fundamentally limited, since these rules are static. The resulting rules will fail after an adversary adapts its techniques. Existing methodologies \cite{xu2024intelex,schwartz2025llmcloudhunter,sigmagen2025,rulegenie2025,falcon2025} are largely focused on a mapping from a report to a rule set, neglecting the necessity for generating rule variants that cover the broader behavioral spectrum of a single attack technique.
%
%
\noindent The right section of Figure~\ref{fig:motivating_example} depicts the approach of \sol. \sol is designed to bridge this gap by transitioning from static rule extraction to a proactive, infrastructure-aware variant generation. Instead of relying on static rules which can be evaded~\cite{canotescape2023} after an adversary performs a technique shift, \sol takes the static rule as input, identifies the possible technique substitutions based on the infrastructure of the environment, and (assisted by LLMs) it generates variant detection rules that cover the technique shifts.

\subsection{Threat Model and Assumptions}
\label{sec:threatmodel}

\subhead{Scope}
We focus on adaptive adversaries who employ known techniques (\eg ATT\&CK patterns) but substitute between them to evade static rules.  We do not attempt to detect entirely novel toolchains or zero-day exploits that do not yet have any established behavioral footprint.  Custom or polymorphic tools that diverge from documented ATT\&CK or CAPEC entries are acknowledged as out-of-scope; addressing those may require complementary runtime or anomaly-based defenses.

\smallskip
\subhead{Assumptions}
We assume the semantic correctness of external knowledge bases (\eg~\mitre and Atomic Red Team~\cite{atomicredteam}), which we treat as ground truth for adversary behavior.
We also assume well-formed Sigma rules from trusted repositories \cite{roth2017sigma}, including accurate technique metadata tags.
Finally, \sol operates offline on infrastructure definitions rather than live telemetry; it does not require streaming production logs and instead relies on static profiles to infer available data sources.

\section{\sol: \underline{A}utomated \underline{R}ule \underline{M}utation \underline{S}olution}
\label{sec:systemarchitecture}

\subsection{System Overview}
\label{sec:system_overview}

\begin{figure}[!t]
    \centering
    \includegraphics[width=\linewidth]{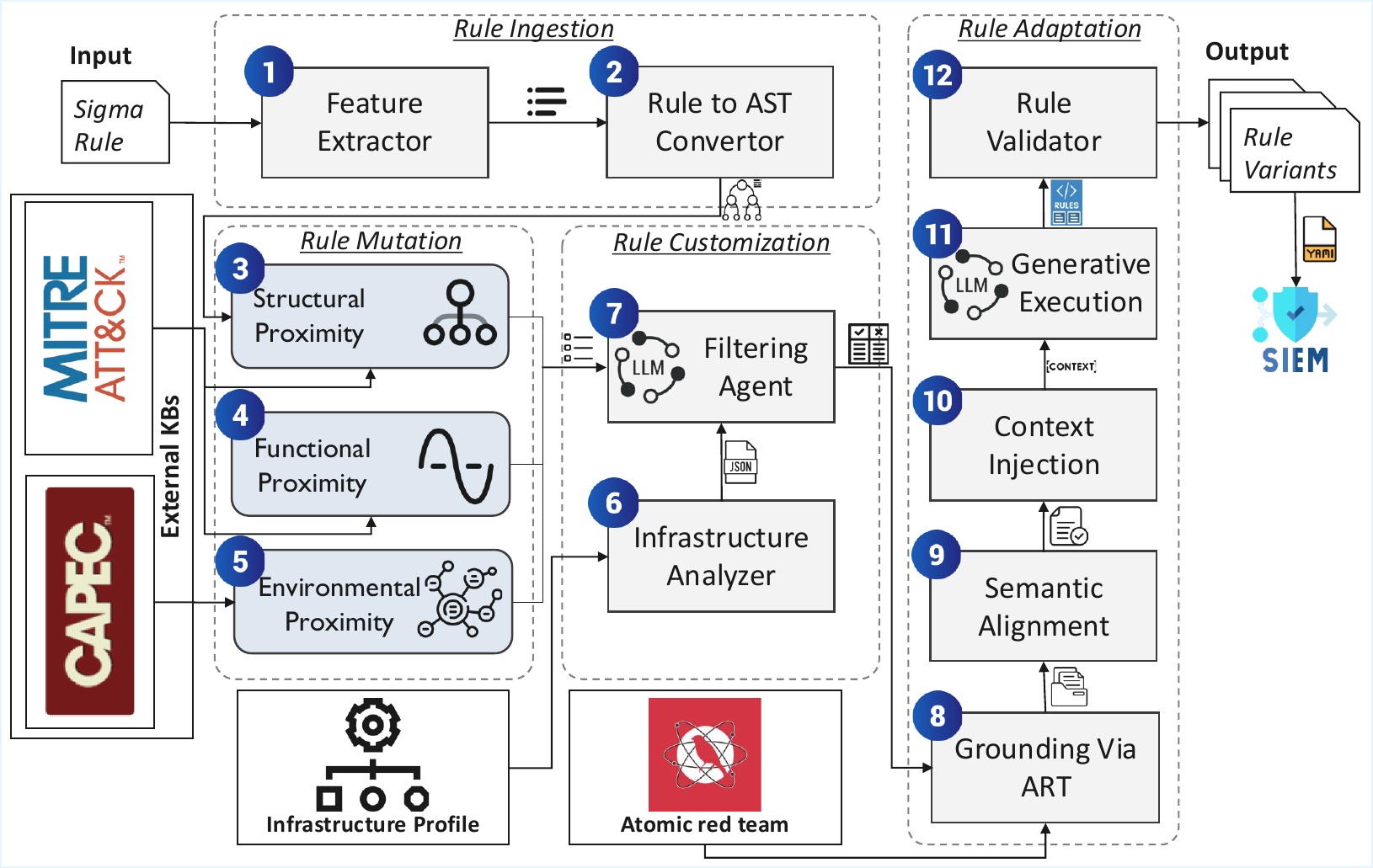}
    \caption{High-level overview of \sol and its components.}
    \label{fig:sol_design}
\end{figure}

Figure~\ref{fig:sol_design} illustrates the architecture of \sol. 
The solution operates in four distinct phases to transform static detection logic into a dynamic, infrastructure-aware defense posture.
The first phase, Rule Ingestion, addresses the syntactic heterogeneity of detection rules. Although Sigma provides a standardized format, individual rule structures often vary. \sol normalizes these inputs by parsing the raw YAML file into a unified Abstract Syntax Tree (AST)~\cite{aho2006compilers}. An AST is a hierarchical tree representation of the syntactic structure of given source code or a rule. Using this method, \sol extracts critical metadata such as detection fields and the underlying attack technique.
The second phase, Rule Mutation, serves as the core engine of \sol and is composed of a three-dimensional proximity logic to identify substitutable techniques. Rather than relying solely on hierarchical relationships, this component searches for semantically equivalent behaviors across different knowledge bases~(\mitre, CAPEC) that fit the structural, functional, and environmental proximity of the rule in question. 
%
The third phase, Rule Customization, acts as a strategic filter between the mutation and adaptation steps. Naively generating rules for every possible technique variant leads to computational waste and alert fatigue if the resulting rules are irrelevant to the target environment. To mitigate this, \sol analyzes the enterprise's specific infrastructure. This agent removes \sepehr{technique} variants that require unavailable telemetry (\eg discarding Linux-specific variants in a Windows-only environment) before they are passed to the adaptation phase.
 
Finally, the fourth phase in \sol is the Rule Adaptation, which contains a RAG-based generative agent responsible for translating the syntax of the original rule to match the newly identified technique variants. It consists of multiple components responsible for grounding the Atomic Red Team based on the techniques, finding the closest test case, constructing the CoT-based prompt, generative execution, and lastly, rule validation. \sepehr{These components help improve reasoning, mitigate hallucinations, and ensure operational validity.}

\subsection{Rule Ingestion}
\label{sub:rule_ingestion}
The initial phase of \sol is the Rule Ingestion engine. Its primary objective is to normalize the heterogeneity of input detection rules into a structured, machine-readable format. This decoupling allows the downstream mutation engines to operate on abstract logic rather than parsing raw text repeatedly. The process is divided into two key components:

\smallskip
\subhead{Feature Extraction}
    \sol extracts the semantic metadata required to understand the adversarial context of the rule. It parses the input rule to identify critical indicators, including the tactics, techniques, and procedures mapped to the \mitre framework, as well as specific Indicators of Compromise such as file hashes, command-line arguments, or registry keys. This information is essential for the subsequent mutation step.

\smallskip
\subhead{Rule to AST Converter}
    \sol converts the rule into an Abstract Syntax Tree. While raw Sigma rules provide a structured format, they are still fundamentally text-based and rigid. Converting them into an AST provides a unified, hierarchical representation of the detection logic (\eg AND/OR conditions, selections, and filters) that is independent of the original syntax. 
    \sepehr{Appendix~\ref{app:ast_example} illustrates an example of a generated AST structure.}

\begin{figure*}[t]
        \centering
        \begin{subfigure}[b]{0.30\textwidth}
            \centering
            \includegraphics[width=\linewidth]{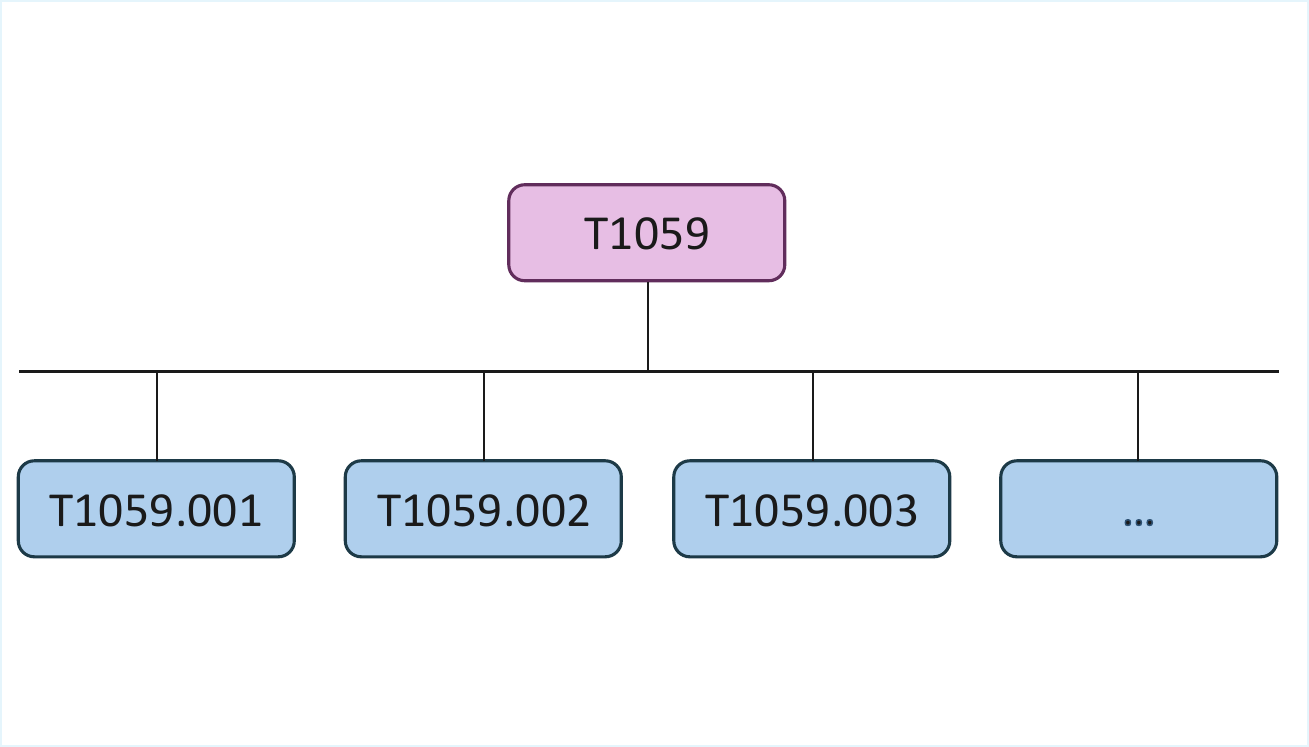}
            \caption{Structural proximity.}
            \label{fig:structural}
        \end{subfigure}
        \hfill
        \begin{subfigure}[b]{0.30\textwidth}
            \centering
            \includegraphics[width=\linewidth]{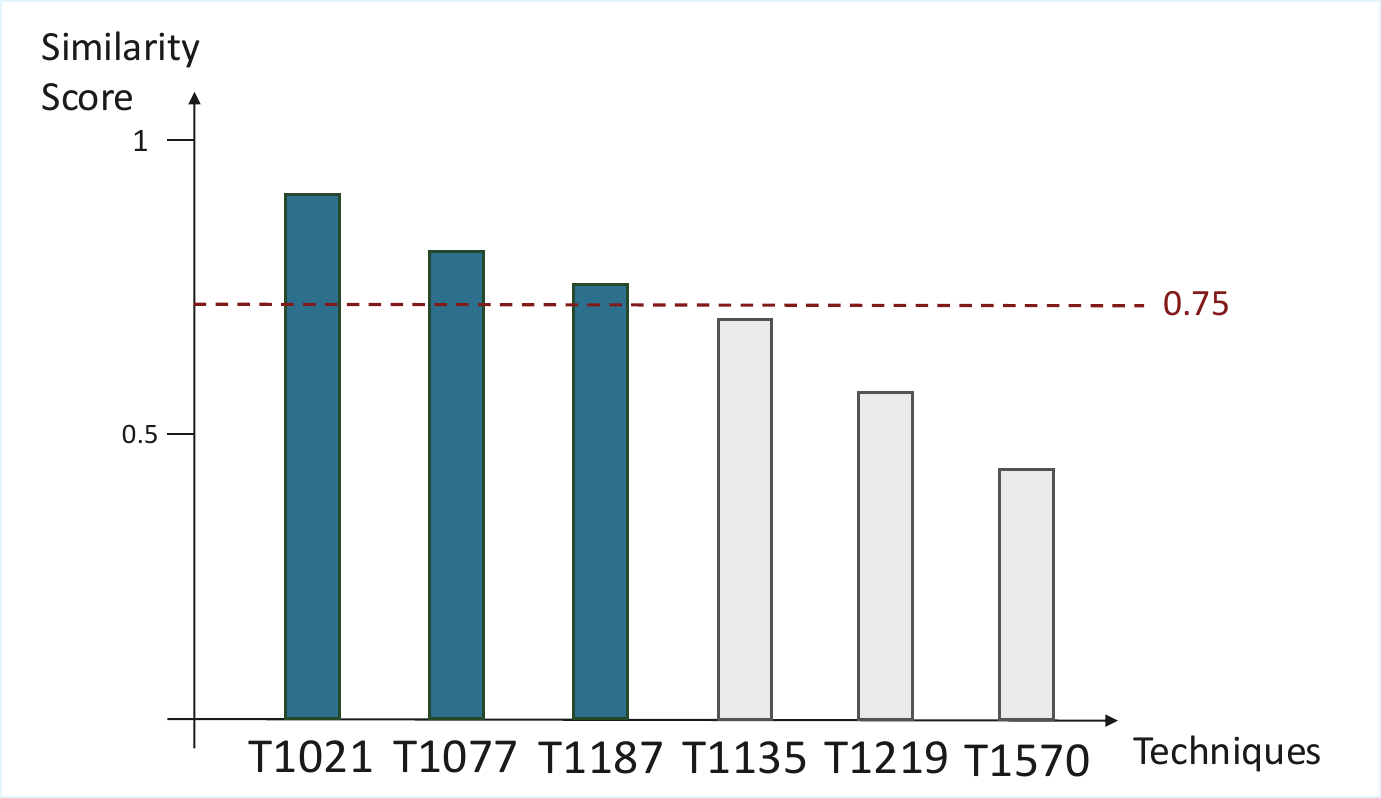}
            \caption{Functional proximity.}
            \label{fig:functional}
        \end{subfigure}
        \hfill
        \begin{subfigure}[b]{0.38\textwidth}
            \centering
            \includegraphics[width=\linewidth]{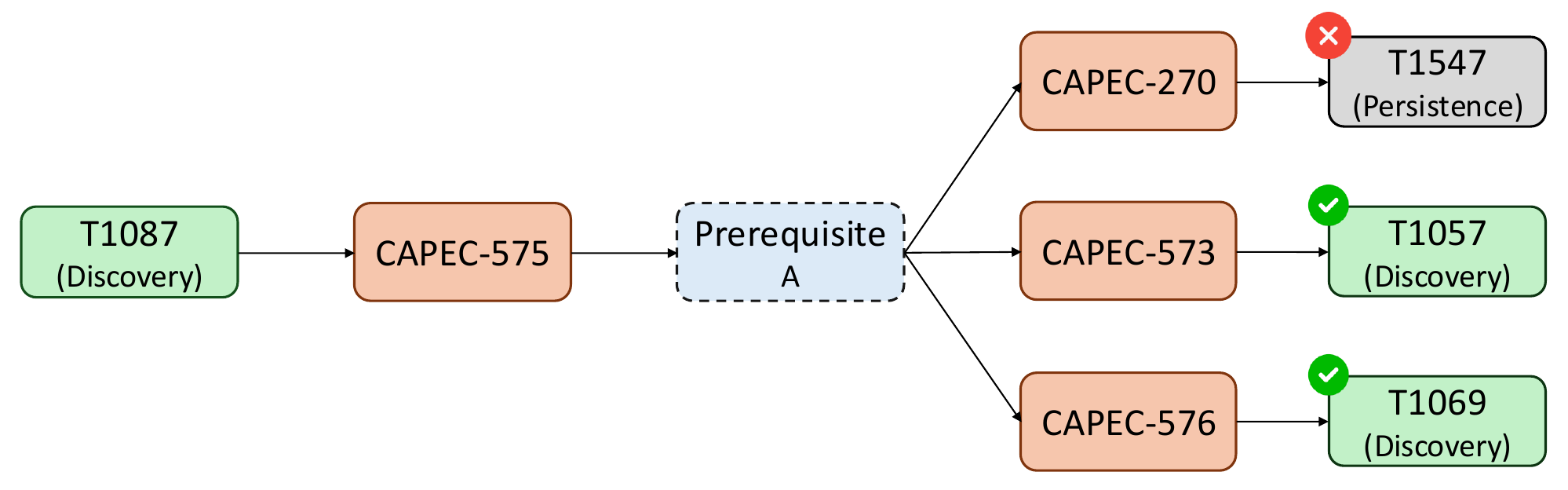}
            \caption{Environmental proximity.}
            \label{fig:environmental}
        \end{subfigure}
        \caption{Example of three-dimensional variant discovery. 
        }
        \label{fig:three_proximities}
    \end{figure*}
    
\subsection{Rule Mutation}
\label{sub:rule_mutation}
The core engine of \sol is the Rule Mutation phase. This phase is responsible for transforming a static seed rule into a diverse set of \sepehr{technique} variants. 
%
\sol identifies alternative attack methods that achieve the same tactical goal as the input rule. It is critical to recognize that a detection rule does not merely represent an alert string; it represents a specific adversarial procedure. Therefore, the first task is to discover interchangeable techniques that can execute the same attack using different mechanisms.
The rule mutation engine identifies variants by analyzing a technique's location within a three-dimensional proximity space:

\smallskip
\subhead{Structural Proximity}
    This dimension leverages the hierarchical taxonomy of the \mitre framework. We define techniques as structurally close if they minimize the shortest path within the \mitre Directed Acyclic Graph (DAG). To construct this DAG, \sol parses the official knowledge base of \mitre, extracts all \texttt{attack-patterns} to instantiate the graph nodes and \texttt{subtechnique-of} relationships to establish directed edges, thus logically linking each sub-technique to its overarching parent or other siblings. Specifically, techniques sharing the same parent technique are considered immediate siblings and highly probable variants.
    Figure~\ref{fig:structural} illustrates this process, where the seed technique {\tt T1059}~(Command and Scripting Interpreter) is expanded into its sub-techniques (\eg~{\tt T1059.001}~PowerShell, {\tt T1059.003} Windows Command Shell), each representing a valid structural variant.
    However, if a technique does not have any parents or siblings, this proximity space will not find any variant techniques. \sepehr{Therefore, there is a need} for other proximity spaces to find relevant techniques. While extending structural proximity to multi-hop techniques (\eg cousins in the DAG) could result in a larger number of variants, it risks introducing semantic drift. \sepehr{To address this}, \sol uses the functional and environmental proximity spaces to safely discover those distant techniques.

\smallskip
\subhead{Functional Proximity}
    While structural proximity finds siblings, it misses techniques that are taxonomically distant but functionally similar. We define functional proximity based on the semantic similarity of technique descriptions, relying on the hypothesis that techniques described with similar language likely serve the same adversarial intent.
    To find this set of techniques, \sol performs Corpus Extraction \cite{zhu2023attackgpt, satvat2021extractor} by harvesting unstructured text descriptions and detection advice from the \mitre knowledge base. We then utilize an embedding model (\eg sentence-transformers model) to project these descriptions into a high-dimensional vector space. \sol calculates the cosine similarity between the original technique vector ($V_{o}$) and all candidate vectors ($V_{t}$), selecting variants that exceed a similarity threshold $\theta$. 
    $
    Sim(V_{o}, V_{t}) = \frac{V_{o} \cdot V_{t}}{\|V_{o}\| \|V_{t}\|} > \theta.
    $
    The threshold was set to 0.75. During preliminary tuning, we saw that thresholds lower than 0.70 resulted in noise and irrelevant techniques, while thresholds above 0.85 severely limited the discovery of valid functional variants.
    Figure~\ref{fig:functional} shows an example where \sol identified {\tt T1021} and 2 other techniques similar to the original technique extracted from the rule.

\smallskip
\subhead{Environmental Proximity}
    It defines closeness based on the system state required to execute an attack. Two techniques are considered isomorphic if they rely on the exact same set of privileges, vulnerabilities, and resources, and they share the same tactic \cite{nour2025threatify}.
    %
    %
    To find this set of techniques, \sol employs cross-framework mapping \cite{mitre2024capec} to link the seed's MITRE ID to its corresponding CAPEC ID (Common Attack Pattern Enumeration and Classification), which contains richer dependency data. We parse fields such as Prerequisites to identify the specific environmental conditions required for the attack. \sol then queries the CAPEC dataset to identify other attack patterns that share identical prerequisites. However, sharing a vulnerability does not automatically imply tactical relevance. To ensure the generated variants serve the same strategic goal as the original attack, we enforce a tactic consistency filter. \sol only accepts techniques that align with the parent tactic of the seed rule, discarding those that rely on the same vulnerability but result in a completely different outcome.
    As shown in Figure~\ref{fig:environmental}, \sol starts with the seed technique {\tt T1087}. It maps this to {\tt CAPEC-575} and extracts its specific prerequisites. Querying the knowledge base for other CAPEC IDs sharing these exact prerequisites yields {\tt CAPEC-270}, {\tt CAPEC-573}, and {\tt CAPEC-576}. These patterns map back to three potential variant techniques: {\tt T1547}, {\tt T1057}, and {\tt T1069}. 
    Finally, \sol applies the tactic-based filtering. Since the original technique {\tt T1087} belongs to the Discovery tactic, the system discards {\tt T1547} as it belongs to the Persistence. {\tt T1057} and {\tt T1069} are retained as valid environmental variants because they satisfy the environmental dependencies while maintaining the original tactical intent.

\subsection{Rule Customization}
\label{sub:rule_Customization}
The third step of \sol is Rule Customization. This phase functions as a strategic filter, positioned between the Rule Mutation and the Rule Adaptation components.
%
Generating rules for techniques that are irrelevant to the organization's specific environment wastes computational resources during the adaptation phase and contributes to alert fatigue in the SOC. The goal of this step is to ensure that \sol only produces rules that are tailored to the enterprise's active infrastructure.

\smallskip
\subhead{Infrastructure Analyzer}
    This component ingests the raw infrastructure profile of the enterprise (\eg ontology or graph-based representation, example in Appendix \ref{app:infra_profile_example}). It performs preprocessing to normalize this data into a structured JSON format, creating a digital representation of the environment's capabilities and constraints (\eg OS types, installed software, logging levels).
    
\smallskip
\subhead{Filtering Agent}
    The core of the Rule Customization phase is an LLM-based filtering agent. Utilizing CoT reasoning, \sepehr{which is a prompting technique that guides the model through step-by-step intermediate logical deductions before reaching a final decision}, this agent evaluates each candidate variant technique against the infrastructure profile. The agent determines applicability based on hard constraints. For example, if the infrastructure profile indicates a Windows-only environment with no Mac OS assets, any variant technique exclusive to Mac OS (\eg~{\tt T1543.004}~Launch Daemons) is immediately discarded. By filtering these out before the computationally expensive adaptation phase, \sol ensures that resources are focused solely on generating high-value rules.

\subsection{Rule Adaptation}
\label{sub:rule_adaptation}
After discovering the relevant technique variants, the final challenge is to adapt the detection rule's syntax to match the new adversarial procedure. A direct syntax translation is often impossible because different techniques (\eg PowerShell vs. WMI vs. Bash) utilize fundamentally different command structures and logging artifacts.
%
Naive heuristic methods ({\tt if-else} logic) \cite{canotescape2023} lack the scalability to handle the infinite variability of command-line obfuscation. Supervised learning approaches \cite{arp2022and} are equally infeasible due to the lack of large-scale, labeled datasets mapping rules to their variants. While zero-shot Generative AI \cite{agarwal2024codemirage} offers flexibility, it is prone to hallucination and lacks the grounded context required for security context.
To address those challenges, \sol employs a RAG Agent enhanced with CoT reasoning \cite{levi2024cyberpalaiempoweringllmsexpertdriven}. This hybrid approach grounds the LLM in verifiable ground truth data while leveraging its reasoning capabilities for syntactic adaptation.

\smallskip
\subhead{Grounding via Atomic Red Team}
    To prevent hallucination, we utilize Atomic Red Team (ART) \cite{atomicredteam}, a repository of scripted cyber attacks mapped to the \mitre framework. We treat the ART repository as our reference for translation. Since a single MITRE Technique ID (\eg~{\tt T1003}) is abstract and may have multiple distinct implementations (\eg using {\tt MimiKatz}, {\tt ProcDump}, or {\tt TaskMgr}), we cannot simply retrieve any or all tests.

\smallskip
\subhead{Semantic Alignment Algorithm}
    To adapt the rule accurately, \sol identifies the specific implementation of the target technique that aligns with the original rule's intent. The selection process is as follows:
    \begin{enumerate*}[label=(\roman*)]
        \item fetches all available Atomic tests associated with the target technique ID;
        \item encodes the original rule's detection logic and description into a high-dimensional vector using an embedding model;
        \item performs a similarity search against the vectorized descriptions of the retrieved Atomic tests; and
        \item identifies the single target test that is semantically closest to the original rule. This ensures we are adapting the rule for the correct atomic test of the new technique.
    \end{enumerate*}
    Appendix~\ref{app:closest_atomic} provides the pseudo code to select the closest Atomic test. 

\smallskip
\subhead{Context Injection}
    Once the target test is selected, we construct a precise translation context containing two ground truths:
    \begin{enumerate*}[label=(\roman*)]
        \item the specific Atomic test command that triggers the original rule (\ie source context), and
        \item the specific Atomic test command for the new variant selected in the target context.
    \end{enumerate*}
    
\smallskip
\subhead{Generative Execution}
    We feed the translation pair to the LLM agent. We instruct the agent to analyze the syntax of the source command, map the underlying detection logic to the target command structure, and craft the Sigma rule entirely. This ensures the new rule detects the behavior of the variant, not just keywords.
    Appendix~\ref{app:prompt_example} illustrates how the prompt is constructed. 
    
\smallskip
\subhead{Rule Validator}
    This is a deterministic module that uses public packages/tools to parse the generated rule to ensure compliance with the Sigma schema, automatically fixing common formatting errors before saving the file for deployment.


\section{Implementation}
\label{sec:implementation}
We built a Proof of Concept of \sol using Python 3.10. 
%
The experiments were conducted on Ubuntu 20.04 with a 64-core processor, 211 GB RAM, and an NVIDIA A100 GPU.

\smallskip
\noindent\textbf{LLM Models:} 
\sepehr{To evaluate \sol, we utilized both local and cloud-based models to account for different privacy and hardware constraints. We used three LLMs: 
\begin{enumerate*}[(i)]
    \item \texttt{Foundation Sec}\cite{foundation_sec_8b}: a fine-tuned version of Meta's Llama 3.1 model for cybersecurity;
    \item \texttt{Lily Cybersecurity}\cite{lily_cybersecurity_7b}: a Mistral fine-tune model with 22000 hand-crafted cybersecurity and hacking-related data pairs; and lastly
    \item \texttt{ChatGPT-4o-mini}\cite{openai_gpt4omini}: through OpenAI APIs.
\end{enumerate*}
To maintain consistency, temperature was set to 0.1 and max tokens to 2000. Furthermore, we included \texttt{ChatGPT-off-the-shelf} (GPT-4o) as our baseline. 
%
We also utilized \texttt{all-mpnet base-v2}\cite{all_mpnet_base_v2} for functional proximity. It is a sentence-transformers model that maps sentences and paragraphs to a 768-dimensional dense vector space, and can be used for tasks like clustering or semantic search.}

\smallskip
\subhead{Public Sigma Rules Repository}
We constructed a seed dataset by randomly selecting 50 verified detection rules from the official SigmaHQ repository~\cite{roth2017sigma}.
To ensure compatibility with \sol's multi-stage pipeline, we required that the MITRE technique associated with each selected input rule have at least one corresponding test case available in the Atomic Red Team library. 
%
\sepehr{To characterize the representativeness of our seed dataset despite this selection constraint and sample size, we stratified the 50 rules across MITRE ATT\&CK tactics and target platforms. The selected rules span 12 of the 14 enterprise tactics, covering Windows (52\%), Linux (36\%), and other platforms (12\%).}

\smallskip
\subhead{Knowledge Base}
\sol incorporates three distinct external knowledge bases to drive its reasoning and mutation engines:
\begin{enumerate*}[(i)]
    \item \mitre knowledge base~\cite{strom2018mitre}, which provides the taxonomy required for our structural and functional variant discovery. We utilized the \mitre Enterprise knowledge base version 14.1, which encompasses 14 tactics, 201 techniques, and 435 sub-techniques;
    \item CAPEC knowledge base~\cite{mitre2024capec} version 3.9, which contains 559 distinct attack patterns. CAPEC provides the rich prerequisite data necessary for the environmental proximity analysis; and
    \item Atomic Red Team repository~\cite{atomicredteam}, which provides concrete, executable command-line tests. The repository contains over 1,300 executable atomic tests that cover more than 300 unique \mitre techniques. 
\end{enumerate*}

\section{Evaluation}
\label{sec:evaluation}
Unlike standard detection engineering tasks~\cite{10251369}, there is no pre-existing dataset or baseline that maps rules to their optimal, infrastructure-aware variants. 
To evaluate \sol, we address three research questions (RQs) designed to assess \sol's output from complementary perspectives.

\rqline{RQ1}{Fidelity: Are the generated rules true behavioral variants of the original attack?}
%
%
%
\noindent This RQ aims to verify if \sol's outputs represent the same underlying adversarial intent rather than unrelated noise or hallucinations.
Measuring the fidelity of security logic is inherently challenging; traditional similarity metrics (such as BLEU \cite{saadany2021bleu} or cosine similarity) are insufficient because they cannot discern whether a syntactic change preserves the rule's detection logic or renders it invalid. Consequently, relying solely on automated similarity scores yields unreliable results.
To address this, we applied cross-validation by:
\begin{enumerate*}[(i)]
    \item asking a security practitioner to evaluate the validity of the generated rules; and
    \item employing an LLM-as-a-Judge: the first LLM analyzes the original seed rule to extract the core adversarial intent, while the second LLM acts as a judge and determines if the variant preserves the original intent while correctly adapting to the new environment.
\end{enumerate*}

\begin{table}[!t]
    \centering
    \caption{Fidelity of rules generated by \sol\ compared with baseline.}
    \label{tab:variant_fidelity}
    \renewcommand{\arraystretch}{1.2}
    \setlength{\tabcolsep}{5pt}
    \arrayrulecolor{gray!90}
    \resizebox{\linewidth}{!}{
    \begin{tabular}{lcccc}
    \rowcolor{gray!15}
    \hline
    \textbf{Metric} &
    \textbf{\sol~Foundation Sec} &
    \textbf{\sol~Lily} &
    \textbf{\sol~ChatGPT} &
    \textbf{ChatGPT-off-the-shelf} \\
    \hline

    Total      & 173 & 179 & 168 & 93 \\
    Valid      & 166 (95.9\%) & 159 (88.8\%) & 154 (91.7\%) & 51 (54.8\%) \\
    Invalid    & 7 (4.1\%) & 20 (11.2\%) & 14 (8.3\%) & 10 (10.8\%) \\
    Unchanged  & 0 (0.0\%) & 0 (0.0\%) & 0 (0.0\%) & 32 (34.4\%) \\
     \bottomrule
    \end{tabular}}
\end{table}

Table~\ref{tab:variant_fidelity} presents the comparison results between \sepehr{\sol with different configurations and ChatGPT-off-the-shelf.} Here, we measured three metrics:
\begin{enumerate*}[(i)]
    \item valid variant: the rule has been successfully mutated and correctly adapted to the new technique/platform while preserving the original intent;
    \item invalid variant: the mutation resulted in broken logic, or a loss of the original detection capability;
    \item unchanged variant: the system failed to trigger a mutation, resulting in a duplicate of the input rule.
\end{enumerate*}
The results show that \sol generated a higher number of total variants (173, 179, 168 vs. 93), owing to the rule mutation component that actively explores alternative techniques.
In terms of quality, \sol demonstrates superior fidelity. \sol achieves a fidelity rate of 95.9\% (166 out of 173 variants), 88.8\% (159 out of 179 variants), and 91.7\% (154 out of 168 variants), respectively under different configurations, compared to only 54.8\% (51 out of 93) for the baseline. This significant margin validates the effectiveness of the rule mutation component in maintaining semantic correctness during mutation.
Notably, \sol achieved an unchanged rate of 0.0\%, whereas the baseline resulted in 34.4\% unchanged outputs. This indicates that ChatGPT-off-the-shelf often struggles to escape the local optima of the original rule, frequently returning the input itself. In contrast, \sol consistently enforces mutation, ensuring that every output was a distinct attempt at adapting the logic to a new technique or platform. 



\rqline{RQ2}{Correctness \& Validation: Are the generated rules semantically valid and convertible to SIEM queries?}
\noindent To assess the operational viability of the generated rules, we decompose the concept of rule correctness into two distinct dimensions: theoretical rule correctness (syntactic validity and semantic logic) and practical end-to-end validation.
%


A rule must not only be syntactically correct to function within a SIEM environment, but must also possess sound internal logic to perform accurate detection.

\begin{figure}[!t]
        \centering
        \includegraphics[width=.8\linewidth]{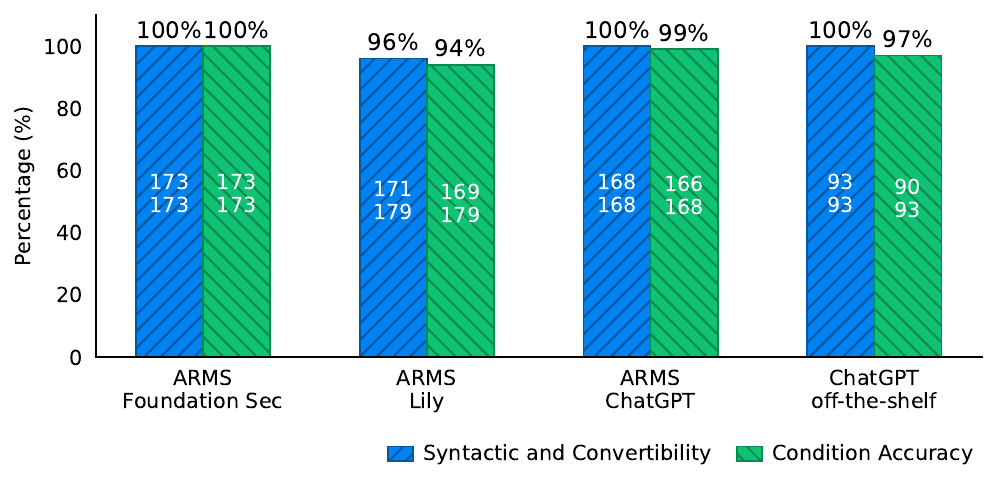}
        \captionof{figure}{Correctness of rules generated by \sol compared with baseline.}
        \label{fig:rule_correctness}
    \end{figure}

\smallskip
\subhead{Syntactic Validity and Convertibility}
    A rule that fails to convert into a target SIEM format is a failure, regardless of its theoretical quality. To evaluate this, we define the rule convertibility as the percentage of generated Sigma rules that can be successfully converted into Splunk and ELK queries without triggering parsing errors or syntax warnings. We utilized the official {\tt Sigma-cli}\cite{sigmacli} tool to attempt this conversion for every generated variant. 
    Figure~\ref{fig:rule_correctness} shows that \sol achieves a rule convertibility rate of 95.5\% while using Lily LLM and 100\% when using Foundation-Sec and OpenAI LLM, successfully converting the generated variants into valid SIEM queries. The perfect conversion score of \sol can be attributed to the structured enforcement of the Sigma schema within the Rule adaptation phase, which first uses CoT reasoning to construct the rule, and second, ensures that all outputs strictly adhere to the required YAML format before they are finalized.

    Beyond simple compilation, the internal logic of the rule must be operationally sound. A rule might be syntactically perfect but operationally useless if it contains broken boolean logic, such as impossible {\tt AND} conditions or disconnected selection fields. 
    To measure this, we adopt the condition accuracy metric established in~\cite{schwartz2025llmcloudhunter}. This metric specifically evaluates the condition field of the Sigma rule to determine if the boolean operators ({\tt AND}, {\tt OR}, {\tt NOT}) correctly connect the detection artifacts to form a coherent query.
    Figure~\ref{fig:rule_correctness} shows that \sol also achieves a minimum condition accuracy of 94.4\%, indicating that most generated variants possessed logically coherent conditions, except in rare cases where the description was too vague or an incomplete rule was generated.


\smallskip
\noindent \subhead{Rule Validation}
We further conducted an end-to-end validation experiment utilizing MITRE Caldera (an automated adversary emulation framework)~\cite{caldera}. 
We built an APTs testbed~\cite{syed2025comprehensive} simulating a real-world enterprise network. 
The testbed consisted of two virtual machines:
\begin{enumerate*}[(i)]
    \item an Ubuntu Linux machine acting as the Command and Control (C2) server (\ie attacker), and
    \item a Windows 10 machine acting as the research subject and target agent (\ie victim).
\end{enumerate*}
On the offensive side, Caldera orchestrated the emulation, communicating with a Go-based Sandcat agent running on the victim VM. 
For telemetry, the Windows VM was instrumented with \textsc{Sysmon}~\cite{sysmon} to capture deep process-level logging (\eg Event ID 1), and PowerShell Script Block Logging (Event ID 4104) was manually enabled via the Windows Registry to capture the full text of executed commands. 
On the analytical side, we used \textsc{Zircolite}~\cite{zircolite}, a standalone Sigma detection tool, to scan the exported \texttt{.evtx} log files against the generated rule sets.
%
%

\begin{figure}[!t]
    \centering
    \includegraphics[width=.8\linewidth]{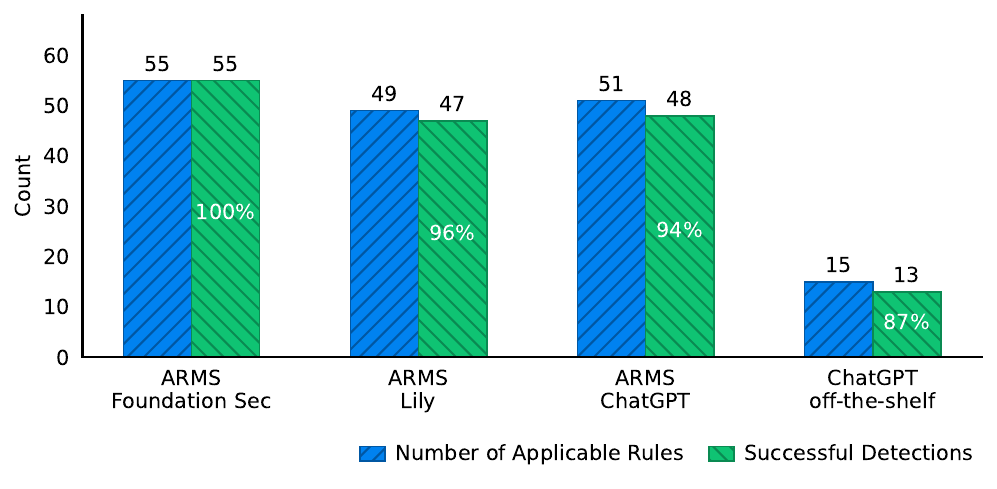}
    \captionof{figure}{Validation of the rules generated by \sol using Caldera~\cite{caldera}.}
    \label{fig:e2e_results}
\end{figure}

Figure~\ref{fig:e2e_results} presents the rule validation results. 
%
All three \sol configurations produced substantially more applicable rules for our setup (49–55) than the baseline (15). This reflects \sol's key advantage: its three-dimensional technique proximity space and infrastructure-aware filtering enable broader, environment-compatible rule expansion, whereas off-the-shelf ChatGPT lacks such environmental awareness and yields far fewer applicable rules.
Moreover, we observe that \sol achieved a highly reliable detection capability, consistently achieving a detection rate of at least 94.1\% across all underlying LLMs. This high success rate directly showcases the efficacy of \sol's rule adaptation phase. By grounding the translation process in verifiable execution commands from the Atomic Red Team repository, the system generates valid, operationally sound rules capable of detecting actual adversarial behavior in a live environment. The baseline, despite generating far fewer rules, still underperformed proportionally with an 86.7\% success rate.

\rqline{RQ3}{Ablation and design choice studies: What is the effect of different components of \sol?}
%
\noindent To evaluate the contribution of individual components within the \sol architecture, we conducted an ablation study focusing on two critical phases: the three-dimensional proximity space and the Rule Customization process. This analysis aims to quantify the impact of each proximity space and the Rule Customization module on the total volume, fidelity, and relevance of the generated rules. 

\begin{figure}[!t]
    \centering
    \includegraphics[width=.8\linewidth]{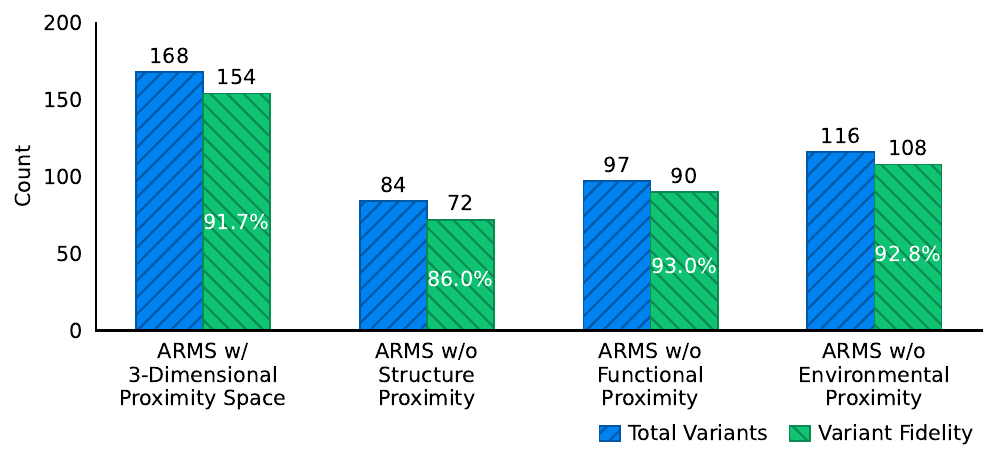}
    \captionof{figure}{Impact of three-dimensional proximity space on the rule fidelity.}
    \label{fig:ablation_fidelity}
\end{figure}

\smallskip
\subhead{Impact of three-dimensional proximity space}
We compared the full \sol against configurations where the structural, functional, or environmental proximity spaces were individually removed. The goal was to determine how each knowledge graph component contributes to the generation of valid attack variants.
As presented in Figure~\ref{fig:ablation_fidelity}, the removal of any single proximity space results in a measurable decrease in the total number of generated variants, confirming that each component actively expands the search space for potential threats. The most significant impact is observed when the structural proximity space is removed; the total number of variants drops by exactly half (from 168 to 84). This indicates that structural analysis is the primary driver for identifying technique variants, likely due to its direct mapping of code-level relationships.
The functional proximity space follows as the second most influential component, while the environmental proximity space shows a slightly smaller but still distinct reduction in output. The lesser impact of the environmental component can be attributed to the current limitations of the CAPEC dataset, which does not yet fully cover every possible technique. However, as these external knowledge bases expand, we anticipate the influence of this component to grow. 
The fidelity percentages remain high and relatively stable across all configurations. 

\begin{figure}[!t]
    \centering
    \includegraphics[width=.8\linewidth]{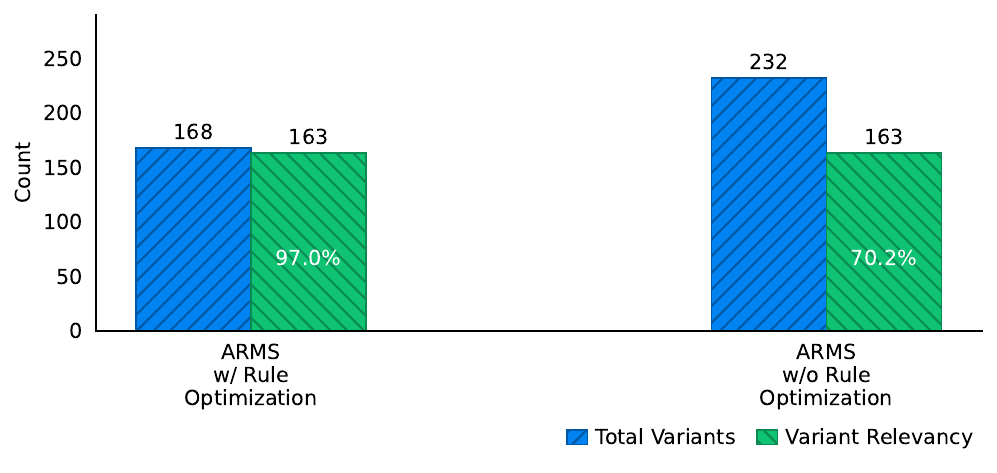}
    \captionof{figure}{Impact of rule customization on rule relevance.}
    \label{fig:ablation_relevancy}
\end{figure}

\smallskip
\subhead{Impact of Rule Customization}
    We compared the full \sol against a version where the LLM-based filtering mechanism was disabled.
    Figure~\ref{fig:ablation_relevancy} highlights the critical role of the Rule Customization phase in aligning generated rules with the target infrastructure. Without the filtering component, the system generated a higher total volume of 232 variants, but the infrastructure relevancy dropped significantly to 70.2\%. This indicates that nearly 30\% of the generated rules were irrelevant to the specific environment, a scenario that would directly contribute to alert fatigue for security operations teams.
    In contrast, \sol with Rule Customization, which incorporates the filtering mechanism, achieved a 97.0\% relevancy rate (163 out of 168 rules). This demonstrates that the Customization phase successfully prunes incompatible or redundant techniques, ensuring that the final output remains highly targeted and actionable for the deployed infrastructure.

\subhead{Functional Proximity Threshold}
    \sepehr{To justify the hyperparameter selection within the functional proximity dimension, we evaluated the system's performance across three distinct cosine similarity thresholds ($\theta$). The functional proximity engine relies on this threshold to determine whether an alternative technique is semantically close enough to the original seed to be considered a viable variant.}
     \sepehr{Table~\ref{tab:threshold_impact} illustrates the trade-off between discovery volume and semantic validity. When applying an overly strict threshold ($\theta = 0.90$), the system behaves too conservatively, generating only 120 total variants. This high threshold severely limits discovery, causing the system to miss 45 valid technique variants that were otherwise successfully captured at the optimal setting. 
    Conversely, lowering the threshold to a permissive state ($\theta = 0.60$) triggers a surge in overall generation, producing 193 total variants. However, the number of valid variants stays the same at 154. This indicates that the additional 25 rules generated at this lower threshold are entirely invalid mutations or hallucinations, which significantly degrades the overall fidelity rate to 79.8\%. Setting the threshold at $0.75$ establishes the optimal outputs, maximizing the discovery of valid attack variants without introducing semantic noise into the mutation.}
    
    \begin{table}[!t]
    \centering
    \caption{Impact of the functional proximity similarity threshold ($\theta$).}
    \label{tab:threshold_impact}
    \renewcommand{\arraystretch}{1.2}
    \setlength{\tabcolsep}{5pt}
    \arrayrulecolor{gray!90}
    \resizebox{.8\linewidth}{!}{
    \begin{tabular}{ccc}
    \rowcolor{gray!15}
    \hline
    \textbf{Threshold ($\theta$)} & \textbf{Total Variants} & \textbf{Variant Fidelity} \\
    \hline
    $0.90$ & 120 & 109 (90.8\%) \\
    $0.75$ & 168 & 154 (91.7\%) \\
    $0.60$ & 193 & 154 (79.8\%) \\
    \bottomrule
    \end{tabular}}
\end{table}
%
   







\section{Discussion}
\label{sec:discussion}

\smallskip
\noindent\textbf{LLM identity mapping:} We observed that off-the-shelf models have a strong tendency towards \textit{identity mapping}, often reproducing the original rule instead of creating a valid variant. By forcing the generation process through our three-dimensional proximity space, \sol effectively breaks this cycle, ensuring every generated rule represents a distinct, meaningful mutation. Furthermore, \sol performs consistently across different LLMs, meaning the choice of model can be dictated by enterprise constraints (\eg inference speed, privacy) rather than model dependence.

\smallskip
\noindent\textbf{Knowledge base dependency:} \sol relies on the completeness of external knowledge bases (CAPEC for environmental proximity, ART for grounding). Techniques without a valid mapping in CAPEC or corresponding Atomic tests cannot currently be mutated. As these repositories expand, coverage will improve; in the meantime, we plan to integrate supplementary sources, such as in-house attack testing frameworks.

\smallskip
\noindent\textbf{Infrastructure profile dependency:} \sepehr{The accuracy of \sol's customization phase depends directly on the completeness of the infrastructure profile, which can be populated via enterprise asset management systems or lightweight discovery scripts. Because environments are inherently dynamic (\eg frequently introducing new SaaS tools, shadow IT, or containerized workloads), an outdated profile leads to inaccurate generation, causing \sol to discard valid variants for unrecorded platforms or retain variants for unsupported platforms. Therefore, the profile must be regularly updated, and \sol re-executed following significant architectural shifts.}

\noindent\textbf{Deployment considerations:} \sepehr{While \sol successfully automates rule mutation, it should not be deployed fully unsupervised immediately. We recommend an initial calibration period with a lightweight human-in-the-loop review gate, where SOC engineers inspect flagged low-confidence mutations (\eg those approaching functional similarity boundaries) prior to deployment to ensure operational safety.}

\section{Related Work}
\label{sec:relatedwork}
%

\subhead{Rule Generation}
Recent work has started to bridge the gap between unstructured sources and detection by generating initial detection rules. 
\cite{schwartz2025llmcloudhunter} uses large language models to generate Sigma rule candidates from cloud-focused open-source intelligence and validates backend conversion success. 
\cite{xu2024intelex} extracts attack-level intelligence from CTI reports and demonstrates downstream Sigma rule generation as part of its evaluation.
Practitioner-oriented systems such as 
\cite{sigmagen2025} aims to reduce the time for rule generation through automated extraction and ATT\&CK mapping. 
While these efforts reduce the cost of producing detections, they largely treat generation as a one-pass task and produce rules that are tightly tied to the described instance. They do not directly address proactive technique substitution, cross-platform behavioral drift, or feasibility under enterprise-specific telemetry constraints. 
%

\smallskip
\subhead{Rule Optimization}
Beyond initial rule creation, recent work has highlighted the operational burden of maintaining large rule sets. 
\cite{rulegenie2025} focuses on SIEM rule set optimization by identifying similar or redundant rules and recommending consolidation and refinement to reduce alert fatigue. This setting is closely related to \sol in that it treats detection rules as the primary artifact and targets real-world operational constraints. However,~\cite{rulegenie2025} optimizes existing rule sets rather than generating behaviorally equivalent rule variants under technique substitution, and it does not explicitly filter outputs according to an enterprise infrastructure profile.
\section{Conclusion}
We proposed \sol, a solution that transforms a single seed Sigma rule into a set of infrastructure-aware behavioral variants. 
\sol models detection engineering as a mutation problem by
\begin{enumerate*}[(i)]
    \item discovering candidate technique substitutions using a three-axis proximity space (\ie structural, functional, and environmental),
    \item filtering candidates using an infrastructure profile to avoid infeasible rules, and 
    \item adapting the seed rule into deployable Sigma variants using a retrieval-augmented generation agent grounded in Atomic Red Team commands.
\end{enumerate*}
We tested \sol using the public SigmaHQ repository. \sol produced between 168 and 179 mutated rules per configuration, with a fidelity rate above 88\%, all syntactically convertible to SIEM with high correctness.

\section*{Acknowledgement}
This work was made possible in part through the support of the National Cybersecurity Consortium and the Government of Canada. It was also supported in part by Ericsson Research and the Security Research Centre of Concordia University.
The authors would like to thank \textit{Jan Willekens} and \textit{Jesus Alatorre} from Ericsson Cyber Defense Center for their invaluable feedback.

\bibliographystyle{IEEEtran}
\bibliography{Ref}

\appendices

\section{Example of the AST generated by \sol}
\label{app:ast_example}
\begin{figure}[!h]
    \centering
    \includegraphics[width=\linewidth]{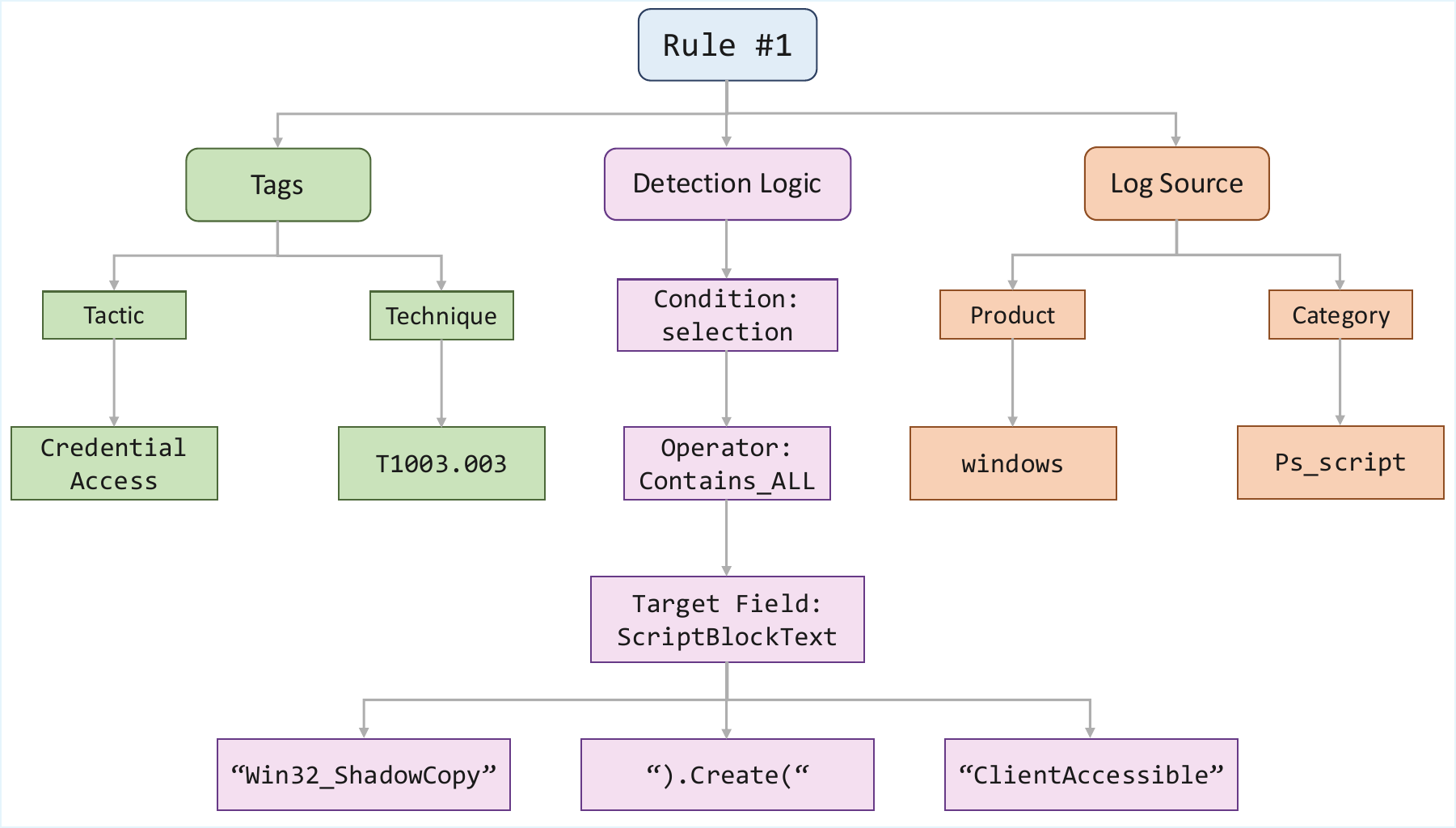}
    \caption{Illustrative example of a generated AST of a Sigma rule.}
    \label{fig:sigma_ast}
\end{figure}

Figure~\ref{fig:sigma_ast} illustrates an example of a generated AST structure of the Sigma rule \texttt{"Create Volume Shadow Copy with Powershell"} from the SigmaHQ repository~\cite{roth2017sigma}. 
The parsing process standardizes the rule into a unified programmatic structure. The resulting AST explicitly isolates the rule's \mitre tactics and techniques and the log source. It captures the detection logic by extracting the target telemetry fields, their associated logical operators (\eg \texttt{Contains\_ALL}), and the specific Indicators of Compromise, all while preserving the original structure of the selection field.

\section{Example of Infrastructure profile}
\label{app:infra_profile_example}
\begin{figure}[!h]
    \centering
    \includegraphics[width=\linewidth]{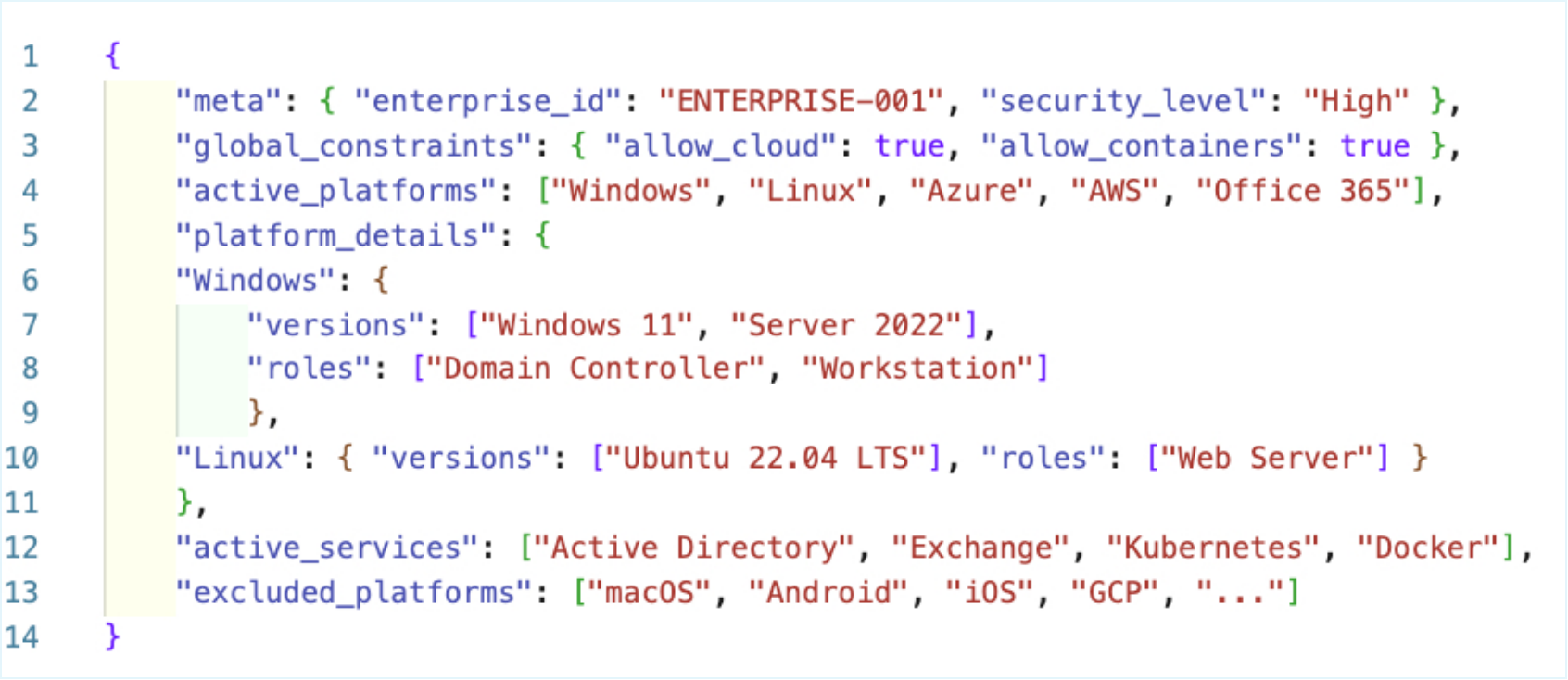}
    \caption{Example of the Infrastructure JSON file.}
    \label{fig:infra_example}
\end{figure}

Figure~\ref{fig:infra_example} illustrates an example of the infrastructure profile JSON file utilized by \sol. This configuration document explicitly defines the approved platforms (Line 4), operating system versions (Lines 7 and 10), and active services present within the target enterprise environment (Line 12), enabling the system to filter out environmentally irrelevant technique variants accurately.

\section{Algorithm for closest Atomic test}
\label{app:closest_atomic}
Algorithm~\ref{alg:closest-atomic-test} outlines the semantic alignment process utilized during the Rule Adaptation phase of \sol. 

\begin{algorithm}[h]
\caption{Select the closest Atomic test for rule adaptation.}
\label{alg:closest-atomic-test}
\begin{algorithmic}[1]
\Require $\tau$: Target technique ID, $r$: original rule, $E$ embedding model 
\Ensure Target Atomic test $t^\star$

\State $T \gets \textsc{FetchAtomicTests}(\tau)$
\State $v_r \gets E(r)$
\ForAll{($t \in T$)}
    \State $v_t \gets E(\textsc{Description}(t))$
    \State $s_t \gets \textsc{CosineSim}(v_r, v_t)$
\EndFor
\State $t^\star \gets \arg\max_{t \in T} s_t$
\State \Return $t^\star$
\end{algorithmic}
\end{algorithm}

\section{Example of prompts}
\label{app:prompt_example}
\begin{figure}[!h]
    \centering
    \includegraphics[width=\linewidth]{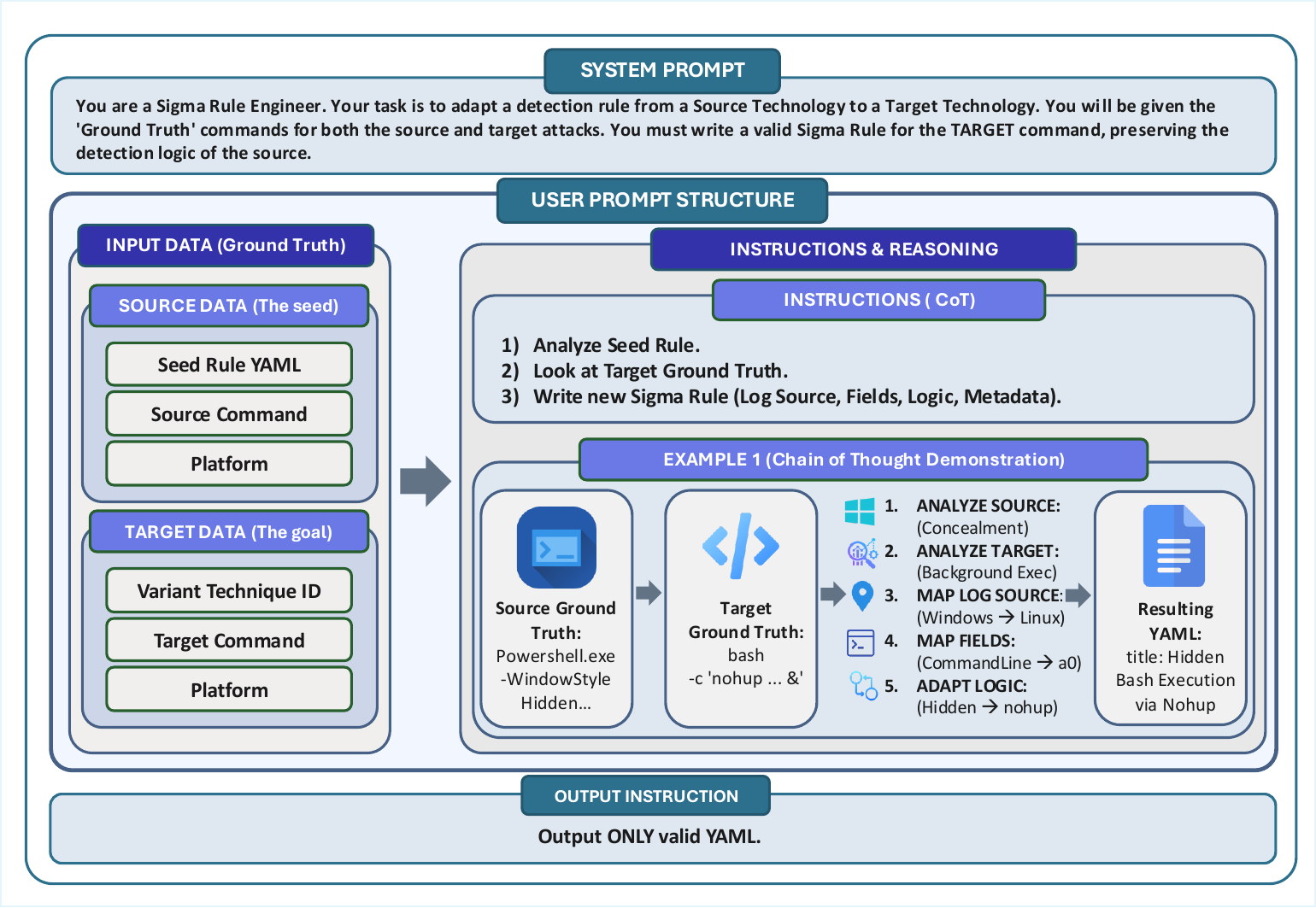}
    \caption{Overview of the prompt given to the RAG-based Adaptor Agent.}
    \label{fig:prompt_adaptor}
\end{figure}

Figure~\ref{fig:prompt_adaptor} shows the prompt setup for the rule adaptation phase. Initially, there is a system prompt that clearly specifies the task the model is expected to perform, which helps prevent hallucinations.
Next, the user prompt section is presented, containing the ground truth data for both the original and target techniques derived from ART data. This section also includes instructions and reasoning steps. The instructions break down the task into several steps, and through a few-shot learning approach, an example of the CoT reasoning is provided. This ensures the model understands the steps to follow and can iterate through a complete generative process.
Finally, the output instruction specifies the desired format for the model's final output.
This comprehensive process is designed to ensure the model delivers the best possible results.

\end{document}